\documentclass[twocolumn,showpacs,showkeys,preprintnumbers,amsmath,amssymb,superscriptaddress]{revtex4}

\usepackage{graphicx}
\usepackage{dcolumn}
\usepackage{bm}

\begin{document}

\title{Morphology of Condon Domains Phase in Plate-Like Sample }

\author{Nathan Logoboy}

\email{logoboy@phys.huji.ac.il}

\affiliation{Grenoble High Magnetic Field Laboratory, MPI-FKF and
CNRS P.O. 166X, F-38042 Grenoble Cedex 9, France}

\affiliation {The Racah Institute of Physics, The Hebrew University
of Jerusalem, 91904 Jerusalem, Israel}

\author{Walter Joss}
\affiliation{Grenoble High Magnetic Field Laboratory, MPI-FKF and
CNRS P.O. 166X, F-38042 Grenoble Cedex 9, France}

\affiliation {Universit$\acute{e}$ Joseph Fourier, B.P. 53, F-38041
Grenoble Cedex 9, France}

\date{\today}

\begin{abstract}

Based on Shoenberg assumption of magnetic flux density dependence of
diamagnetic moments which accounts for an instability of strongly
correlated electron gas at the conditions of dHvA effect and
diamagnetic phase transition (DPT) to non-uniform phase, we
investigate the morphology of the Condon domains (CD) in plate-like
sample theoretically. At one period of dHvA oscillations the
intrinsic structure of inhomogeneous diamagnetic phase (IDP) is
governed by the first order phase transitions between different
non-uniform phases similar to the high-anisotropy magnetic systems
of spin origin, and strongly affected by temperature, magnetic field
and impurity of the sample due to the electron correlations. The
phase diagrams of evolution of IDP with temperature and small-scale
magnetic field in every period of dHvA oscillations are calculated.

\end{abstract}



\maketitle

\section{\label{sec:Introduction}Introduction}

Strongly correlated electron systems are subject of constant
interest of physical community. The instability of an electron gas
due to strong electron correlations at the conditions of dHvA effect
resulting in diamagnetic phase transition (DPT) into inhomogeneous
diamagnetic phase (IDP) with formation of Condon domains (CDs)
\cite{Condon}, \cite{Condon_Walstedt} is intensively studied both
theoretically and experimentally \cite{Kramer1}-\cite{Logoboy4}. The
realization of intrinsic structure of IDP is governed by the
competition between long-range dipole-dipole interaction and
short-range interaction related to the positive interface energy
with typical magnetic length of Larmor radius $r_{c}$ \cite{Condon},
\cite{Privorotskii}.

There are striking similarities between IDP in normal metals and
other strongly correlated systems which undergo phase transition on
temperature and magnetic field with formation of complex macroscopic
patterns, e. g. the type-I superconductors and thin magnetic films
with quality factor exceeding unity. The different technics,
including the powder pattern and magneto-optic methods, successfully
used for observation of intermediate state of type-I superconductors
revealed a very rich structure \cite{Livingstone} which in spite of
all it complexity, amazingly reminds the variety of domain
structures in thin film \cite{Hubert}. The list of the phenomena can
be further continued by including other strongly correlated systems
which driven by the corresponding parameters exhibit the analogous
many-pattern behavior, e. g. Quantum Hall effect system, showing the
charge density wave instability with formation of stripe and bubble
phases \cite{Koulakov}, spin-glasses \cite{Mezard}, itinerant
metamagnets \cite{Yamada}-\cite{Binz}, the Saffman-Taylor
instability of fluid-ferrofluid interface in rotating Hele-Shaw
cells \cite{Miranda} and Langmuir monolayers \cite{Heinig}. At the
background of these phenomena, the model of simple lamina periodic
domain structure (PDS) of normal metals
\cite{Shoenberg}-\cite{Itskovsky} associated with CDs and used for
explanation of electron instability seems to be rather simplified.
Indeed, the new experiments on observation of diamagnetic domains by
use of a set of Hall probes at the surface of the plate-like sample
of silver \cite{Kramer1} reveal rather complicated domain structure
which only at certain conditions, with tilting the sample relative
to applied field, transforms into regular laminar structure similar
to the observation of intermediate state of type-I superconductor
\cite{Sharvin}. Muon spin rotation spectroscopy ($\mu$SR) being a
powerful method for investigation of complex materials was
successfully used for studying CDs \cite{Solt1}-\cite{Solt4}. The
series of excellent experiments on investigation of diamagnetic
instability by methods of $\mu$SR spectroscopy
\cite{Solt1}-\cite{Solt4} prove the existence of CD phase in
beryllium, white tin, aluminum, lead and indium, but reveal also
contradictions in attempts of quantitative analysis in the framework
of existent theoretical considerations.

The specific case of evolution of domain structure for
two-dimensional electron gas in ultra-thin magnetic films with
finite thickness $L \lesssim r_{c}$ was considered in
\cite{Markiewicz}. It was shown that the electron system undergoes
the phase transition into a domain state which can cross over into a
modulated, or vortex, state with decreasing of the sample thickness.
The consideration was restricted by the center of the period of dHvA
oscillations. While it is known that IDP occupies only a part of
dHvA period, it would be extremely interesting to investigate the
morphology of the domain patterns with the change of the magnetic
field within one dHvA period. The influence of magnetic field on PDS
at the conditions of diamagnetic instability was investigated in
\cite{Itskovsky} where the dependencies of the period of domain
structure on magnetic field and film thickness were calculated
numerically. In particular, it was demonstrated by numerical
calculations that the width of the domains with favorable direction
of magnetization increases and the width of the domains with
unfavorable direction of magnetization decreases with field goes
away from the center of dHvA period. The transformation of PDS into
modulated state near the critical point $a=$1 at the center of dHvA
period was also illustrated. Unfortunately, the critical fields
corresponding to transitions between different domain structures
within dHvA period, including bubble lattice, separated bubbles and
so on were not calculated and the phase diagrams were not
constructed.

In \cite{Logoboy4} the equilibrium set for the system of strongly
correlated electron gas at the conditions of dHvA effect was
investigated in the framework of catastrophe theory. It was shown
that in every period of dHvA oscillations the discontinuities of
order parameter accompanied DPT is handled by Riemann-Hugoniot
catastrophe implying the standard scenario for DPT, e.g. DPT is of
the second order at the center of dHvA period, weakly first order in
the nearest vicinity of this point and is of the first order at the
rest part of the dHvA period. Thus, similar to other magnetic
systems, e. g. the spin \cite{Hubert} and metamagnetic ones
\cite{Binz}, the DPT can be realized in rich background of
nonuniform phases depending on the shape of the sample. The studies
of diamagnetic instability \cite{Logoboy4} are based on postulates
of the phase theory when the irreversibility effects are ignored,
and the results are applied for ellipsoidal uniformly magnetized
phases in uniform external magnetic field. Thus, the shape effect
accounted for by use of the conception of demagnetizing coefficient
$n$ consists in expansion of the magnetic field range for IDP
existence $\mid x \mid \le x_{c}(n)$ with increase of $n$.

The calculations based on bifurcation theory allows to eliminate all
possible stable equilibria of the strongly correlated electron
system, but the conclusion about realization of non-uniform states
and evolution of IDP can be done only on the basis of consideration
of global minimum of the free energy of a system for given
configuration of domain patterns, including the energy of long-range
dipole-dipole interaction, the energy of the interphase boundaries
and the magnetic energy in applied field. The calculations become a
rather complicated problem due to existence of non-local
dipole-dipole interaction which can not be accounted for by use of
demagnetizing coefficient contrary to the studies performed in the
framework of phase theory. Although, the relevant problems were
solved in the physics of magnetism of spin origin \cite{Hubert}, the
general picture of formation of non-uniform phases is complicated by
the electron correlations due to non-local magnetic interaction
between electrons \cite{Shoenberg}.

Using the methods developed in the physics of magnetic materials,
for a plate-like sample we calculate the phase diagrams related to
complicated intrinsic structure of IDP which can realized through
the formation of variety of different domain patterns, including
modulated states in the nearest vicinity of critical point and
regular parallel band domains which transferred into separated band
domains, close-packed bubble lattices and isolated bubbles with
changing the small-scale magnetic field $x$ within one period of
dHvA oscillations. We show that evolution of domain structure with
temperature is different at the center of the period of dHvA
oscillations and its other part within the range of existence of
IDP.

The paper is organized as follows. In Sec.~\ref{sec:Model}, we
introduce the model and relevant results of catastrophe theory which
at proper choice of control variables provides a convenient tool for
investigation of DPT. In Sec.~\ref{sec:Results and Discussions} we
present and discuss the general conditions for realization of the
DPT in the sample with taking into account the long-range
dipole-dipole interaction and calculate the critical fields which
characterized the non-uniform phases. We show that convectional
scenario of transformation of magnetic phases is influenced by
electron correlations resulting in strong temperature and magnetic
field dependency of critical fields. Finally, in
Sec.~\ref{sec:Conclusions}, we summarize our main conclusions.

\begin{figure} [t]
  \includegraphics[width=0.4\textwidth]{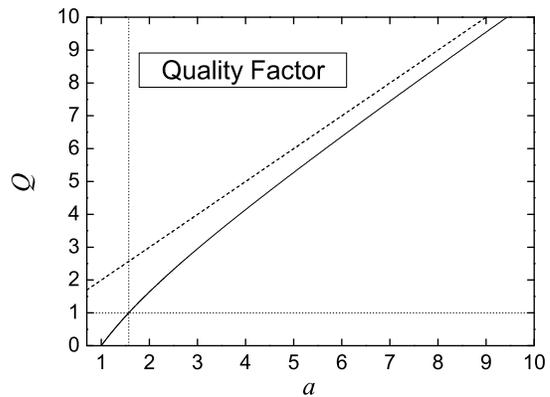}
\caption{\ Quality factor $Q$ Eq.~(\ref{eq:Quality Factor}) defined
for the plate-like sample of normal metal at the conditions of
strong dHvA effect is plotted as a function of reduced amplitude of
dHvA oscillations $a$ (solid line). For $a\ge \pi/2$ the
macrostructure of domain patterns are expected to be similar to the
domain structure of thin high-anisotropy magnetic films of spin
origin due to $Q\ge 1$. The dash line is asymptote of function
$Q=Q(a) \to a+1$ for $a\to +\infty$. } \label{Quality Factor}
\end{figure}

\section{\label{sec:Model}Model}

The properties of strongly correlated electron system at the
conditions of dHvA effect in one harmonic approximation are
described by the free energy functional
\begin{equation} \label{eq:Free Energy Functional 1}
G(y;a,x)=G^{(unif)}+G^{(grad)}, \qquad \quad \\
\end{equation}
where
\begin{eqnarray} \label{eq:Free Energy Functional 2}
G^{(unif)}=a \cos {(x+y)}+\frac{1}{2}y^{2},\qquad  \\ \nonumber
G^{(grad)}=\frac{1}{2}a r^{2}_{c}(\partial_{\zeta} y)^{2}.\qquad
\qquad
\end{eqnarray}
Here, the small-scale magnetic field $x=k\mu_{0}(H-H_{a})$ is the
increment of the large-scale internal magnetic field $\mu_{0}H$ and
external (applied) magnetic field $\mu_{0}H_{a}$, $y$ is oscillating
part of reduced magnetization, $k=2\pi F/(\mu_{0}H_{a})^{2}=2\pi
/\Delta H$, $F$ is the fundamental frequency of the dHvA
oscillations corresponding to the extreme cross-section of Fermi
surface, $\Delta H$ is dHvA period and $a=\mu_{0}\max\{\partial
M/\partial B\}$ is differential magnetic susceptibility
\cite{Shoenberg}. In physical units $x$ is of the order of
$\thicksim$1-10 mT depending on the properties of the electron
system, while $\mu_{0} H$ is $\thicksim$1-10 T. The gradient term in
Eq.~(\ref{eq:Free Energy Functional 2}) accounts for the short-range
correlations on the scale of $r_{c}$ \cite{Privorotskii}, $\zeta$ is
coordinate.

In the framework of bifurcation theory the diamagnetic instability
can be handled by cusp catastrophe \cite{Logoboy4}. In case of thin
film one can show that the bifurcation set in the $a-x$ plane of
control variables $a$ and $x$ is described by the following
expression (see Appendix~\ref{sec:Cusp Catastrophe} for details)
\begin{equation} \label{eq:Bifurcation Set}
a\cos(\sqrt{a^{2}-1}-\mid x_{c}-y(a,x_{c}) \mid)=1,
\end{equation}
where the magnetization $y=y(a,x_{c})$ is given explicitly by
$y=a\sin~(x_{c}+y)$. Eq.~(\ref{eq:Bifurcation Set}) defines the
function $a=a(x_{c})$ which is plotted in Fig.~\ref{Phase
Diagrams}(a). The bifurcation curve $a=a(x_{c})$ divides the $a-x$
plane onto two parts, the inner part is occupied by IDP, while the
outer one corresponds to uniform state. Offering a systematic way to
study the diamagnetic instability, the catastrophe theory implies
the different scenario of formation and evolution of domain patterns
at the center of the period of dHvA oscillations and away of it. At
the center of dHvA period, the stability criterion for CD formation
is $a\ge$1 and theory predict the second order phase transition. The
crossing of bifurcation set with lowing of temperature results in
appearance of two-fold degenerate equilibria, thus, one would expect
the stratification of the sample into the domains of equal widths
and appearance of PDS due to the essential decrease in magnetostatic
energy. The steady-state domain size is defined by competition
between long-range dipole-dipole interaction and short-range
electron interaction on the scale of $r_{c}$ and calculated by
standard method based on minimization of total energy containing two
terms, e. g. dipole-dipole energy energy and surface energy of
separation of two domains \cite{Hubert}. Away of the center of dHvA
period the transition is of the first order and takes place at lower
temperature. At this condition the crossing of bifurcation set leads
to appearance of additional local minimum in free energy. The energy
corresponding to this new minimum is higher than the energy of the
previous state. But, even in this case the formation of non-uniform
phase can decrease the net energy due to decrease of magnetostatic
energy. Thus, at the vicinity of phase boundary a probable new
non-uniform phase consists of fully occupied areas with equal
magnetization and depletion regions with different magnetization,
e.g. separate bubbles embedded into uniformly magnetized sample.
With further decrease of temperature the energy difference between
two local minima in free energy decreases, and further decrease of
net energy can be achieved by organizing the separated bubbles into
bubbled lattice and formation PDS of band domains with volume
fraction linear dependent on magnetic field.

In Sec.~\ref{sec:Results and Discussions} we show that in plate-like
sample the IDP has complicated intrinsic structure and calculate the
phase diagrams corresponding to evolution of IDP within the period
of dHvA oscillations.

\begin{figure}[t]
  \includegraphics[width=0.4\textwidth]{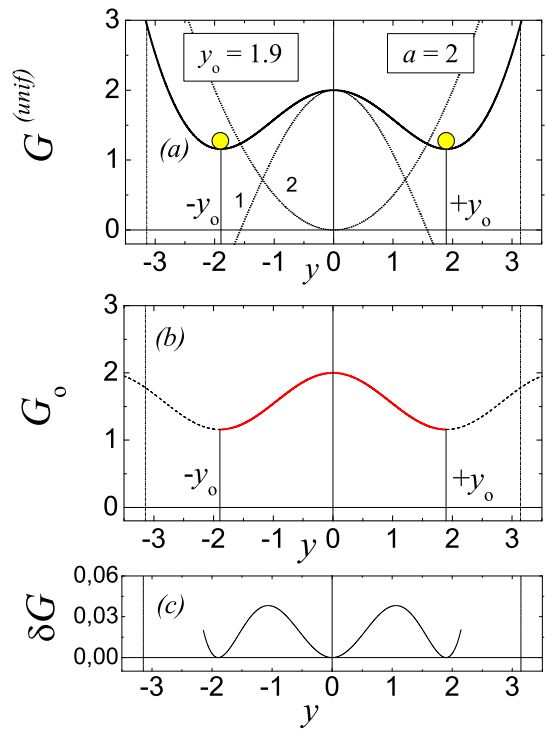}
\caption{\ (color online) Shown are (a) the free energy density
$G^{(unif)}$ (solid line) as a function of magnetization $y$ with
two contributions ($\it curves~1$ and $\it 2$) according to
Eq.~(\ref{eq:Free Energy Functional 2}); (b) the approximated free
energy density $G_{0}$ Eq.~(\ref{eq:Approximation}) at the relevant
range $y \le \mid y_{0} \mid$ where $\pm y_{0}=\pm$1.9 are two
minima and (c) the free energy difference $\delta
G=G^{(unif)}-G_{0}$. The accuracy of the approximation
$G^{(unif)}\approx G_{0}$ is $\sim $10$^{-2}$. The calculations are
made at the center of the period of dHvA oscillations with the value
of differential magnetic susceptibility $a=$2.}
\label{Approximation}
\end{figure}

\section{\label{sec:Results and Discussions}Results and Discussions}

Before proceeding with calculation we define the quality factor for
a plate-like sample according to
\begin{equation} \label{eq:Quality Factor}
Q=\frac{ \partial^{2}_{yy} G^{(unif)}}{\partial^{2}_{yy} G^{(d-d)}}\Bigg\arrowvert_{y=y_{0}}=1-a\cos{y_{0}},  \\
\end{equation}
where $G^{(d-d)}=y^{2}_{0}/2$ is demagnetizing energy for uniformly
magnetized plate-like sample. Here, $y_{0}=y_{0}(a)$ is the
magnetization of two-fold degenerate uniform ground state given
explicitly by the equation
\begin{equation} \label{eq:Uniform Magnetization}
y_{0}-a \sin y_{0}=0
\end{equation}
at $a\ge$1. The function $Q=Q(a)$ is plotted in Fig.~\ref{Quality
Factor} which shows that $Q \ge$1 for all values of $a \ge \pi/$2.
Differential magnetic susceptibility $a=a(\mu_{0}H,T,T_{D})$ is a
function of magnetic field $\mu_{0}H$, temperature $T$ and Dingle
temperature $T_{D}$. Therefore, it follows from Fig.~\ref{Quality
Factor} that in a wide range of temperature and magnetic field
corresponding to the diamagnetic instability, the magnetic
properties of strongly correlated electron gas at the conditions of
the diamagnetic instability are expected to be analogous to the
conventional high-anisotropic magnetic materials which are
characterized by quality factor $Q >$1.

Variation of the free energy $G$ Eq.~(\ref{eq:Free Energy Functional
1}) with respect to the magnetization $y$ at the center of dHvA
period leads to the following differential equation
\begin{equation} \label{eq:DE for Magnetization}
a \sin {y}-y+a r^{2}_{c} \partial^{2}_{\zeta \zeta} y=0 \\
\end{equation}
The first integral of Eq.~(\ref{eq:DE for Magnetization})
\begin{equation} \label{eq:First Integral}
a r^{2}_{c}(\partial_\zeta y)^{2}=2 a \cos{y}+y^{2}+ C, \\
\end{equation}
forms the basis for investigation of non-uniform phases in
one-dimensional problems. By choosing a different integration
constant $C$ one can obtain the solutions describing separate domain
walls, PDS and modulated structures. The corresponding problems in
physics of magnetic materials were solved long ago (see,
\cite{Hubert}). We find it useful to present the analytical
solutions of Eq.~(\ref{eq:DE for Magnetization}) related to PDS for
the system exhibiting the diamagnetic instability, the first, to see
the close analogy with the physics of magnetic phenomena of spin
origin and, the second, the periodic analytical solutions for IDP
were not reported as far as we aware of.

\begin{figure}
  \includegraphics[width=0.4\textwidth]{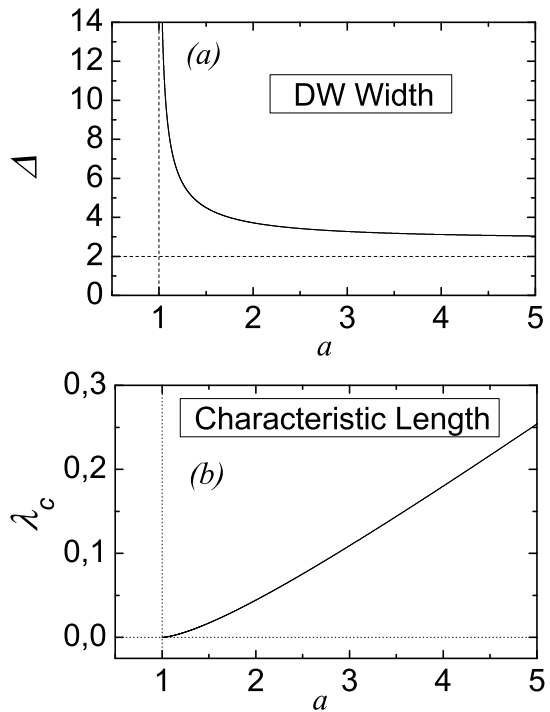}
\caption{ (a) Domain wall width $\Delta$ and (b) characteristic
length $\lambda_{c}$ are plotted as functions of reduced amplitude
of dHvA oscillations $a$. The dash lines in (a) show the asymptotes.
In calculations of $\lambda_{c}$ (b) the value of $r_{c}/L=10^{-2}$
was used.}\label{DW Characteristics}
\end{figure}

In a limit $a-1 \to $0$^{+}$, one can use expansion of RHS of
Eq.~(\ref{eq:DE for Magnetization}) in powers of $b$ due to $b\le
b_{0}=\sqrt{6(a-1)}<<$1. Thus, Eq.~(\ref{eq:DE for Magnetization})
can be integrated resulting in periodic distribution of magnetic
induction $b=b_{0}~\mathrm {sn} (\zeta/\delta_{1},k)$ with a period
$D$=4$K(k)\delta_{1}$, where $\mathrm {sn}$ is Jacobi elliptic
function, $K$ is complete elliptic integral and $k$ is elliptic
modulus. The parameter $\delta_{1}=r_{c}\sqrt{(1+k^{2})/(a-1)}$ is
the domain wall width. At $k=$1, $K \to \infty$ we arrive to the
solution $b=b_{0}\tanh(\zeta/\delta_{1})$ describing the separate
domain wall \cite{Markiewicz}. At $k=$0, we obtain the modulated
structure $b=A \sin(q_{0}\zeta)$ \cite{Markiewicz}, where
$A=\sqrt{6(a-1)}$ and $q_{0}=r^{-1}_{c}\sqrt{a-1}$ are the amplitude
and wave vector.

In case of $a-1 \gtrsim 1$, one can see that the main contribution
is due to the first term in $G^{(unif)}$. In this case the quality
factor $Q>$1 (see, Fig.~\ref{Quality Factor}), the sample is
characterized by high anisotropy, and at the relevant range $y\le
\mid y_{0}\mid$ with a great accuracy the potential $G^{(unif)}$ can
be approximated by trigonometric function (Fig.~\ref{Approximation})
corresponding to two-fold degenerate equilibria $y=\pm y_{0}$
\begin{equation} \label{eq:Approximation}
G^{(unif)}\approx
G_{0}=-\frac{1}{2}K_{0}\sin^{2}(\frac{\pi}{2}\frac{ y}{y_{0}}),
\end{equation}
where
\begin{equation} \label{eq:Anisotropy Constant}
K_{0}=4a\sin^{2}\frac{y_{0}}{2}(1-a\cos^{2}\frac{y_{0}}{2})> 0
\end{equation}
is a function of $a$ and $y_{0}$ and analogous to the constant of
easy-axis crystallographic anisotropy in physics of magnetic
materials. In Eq.~(\ref{eq:Approximation}) we omitted the
unessential constant. Introducing a new variable $\Theta=\pi
y/2y_{0}$, we rewrite Eq.~(\ref{eq:Free Energy Functional 1}) in a
form
\begin{equation} \label{eq:Free Energy Functional 3}
G=-\frac{1}{2}K_{0}\sin^{2}\Theta + \frac{1}{2}A (\partial_{\zeta}
\Theta)^{2},
\end{equation}
with account of Eq.~(\ref{eq:Approximation}). The parameter
$A=a(2r_{c}y_{0}/\pi)^{2}$ characterizing the the short-range
interaction is analogous to the exchange constant for spin systems.
This confirms the close analogy between high-anisotropy
ferromagnetic and the system exhibiting the diamagnetic instability.
Minimization of free energy density Eq.~(\ref{eq:Free Energy
Functional 3}) allows to obtain the non-uniform state described by
periodic function
\begin{equation} \label{eq:PDS}
y=\frac {2y_{0}}{\pi} \sin^{-1}[\mathrm {sn}
(\frac{\zeta}{\delta_{2}},k)],
\end{equation}
where $\delta_{2}=k\sqrt {A/K_{0}}$. If $a$ much greater than 1,
then it is follows from Eq.~(\ref{eq:Uniform Magnetization}) that
$y_{0}\approx \pi$, the parabolic term in Eq.~(\ref{eq:Free Energy
Functional 2}) is negligible, and the solution has a simple form
$y=2 \sin^{-1}\mathrm {sn} (\zeta/\delta_{2},k)$ with
$\delta_{2}=r_{c}k$. At $k=$1, $K \to \infty$ we arrive to the
solution $ y=2\sin^{-1}\tanh(\zeta/\delta_{2})$ analogous to
separate 180$^{0}$ Bloch domain wall.

Similar to the high-anisotropy magnetic films in perpendicular
magnetic field a variety of different domain patterns can form in a
thin film of normal metal at the condition of diamagnetic
instability $a \ge a_{c}$ where $a_{c}$ depends on small-scale
magnetic field $x$ and is defined explicitly by
Eq.~(\ref{eq:Bifurcation Set}). Following \cite{Kooy}, \cite{Cape}
and \cite{Hubert} we calculated the phase diagrams in the $a-x$
plane describing the intrinsic properties of IDP at one period of
dHvA oscillations. In calculations we neglect the structure of the
domain walls assuming that the size of the domain is large compared
to the domain wall width, thus, characterizing the domain wall by
specific wall energy
\begin{equation} \label{eq:Specific DW Energy}
\sigma=\frac{4r_{c}}{\pi}a^{1/2} y_{0}[Q^{2}-(a-1)^{2}]^{1/2}
\end{equation}
and dimensionless domain wall width
\begin{equation} \label{eq: DW Width}
\Delta=\frac{\delta}{ r_{c}}=\frac{4}{\pi} \frac{a^{1/2}
y_{0}}{[Q^{2}-(a-1)^{2}]^{1/2}}. \quad
\end{equation}
We introduce also the commonly used in phase diagram calculations
\cite{Hubert} dimensionless characteristic length
\begin{equation} \label{eq: Characteristic Length}
\lambda_{c}=\frac{\sigma}{2L}, \qquad \qquad
\end{equation}
where $L$ is the width of the plate. The qualities
$\sigma=\sigma(a)$ Eq.~(\ref{eq:Specific DW Energy}),
$\Delta=\Delta(a)$ Eq.~(\ref{eq: DW Width} and
$\lambda_{c}=\lambda_{c}(a)$ Eq.~(\ref{eq: Characteristic Length}
are characterized by strong dependencies on temperature, magnetic
field and purity of the sample through the differential magnetic
susceptibility $a$ (Fig.~(\ref{DW Characteristics}). At the vicinity
of point $a =1$, the behavior of $\Delta$ and $\lambda_{c}$
signalizes about the presence of critical point of the system. Near
the critical point, $a \to 1+0^{+}$, the spontaneous fluctuations
become large, the length scale of fluctuations $\xi \sim \delta \sim
(a-1)^{-1/2}$ has a power law of divergence, and the characteristic
length $\lambda_{c} \sim \sigma \sim (a-1)^{3/2} \to $0. There is no
typical scale length except of the trivial lower (atomic distance)
and upper macroscopic (the size of the system $L$) size scales.

\begin{figure}[b]
  \includegraphics[width=0.4\textwidth]{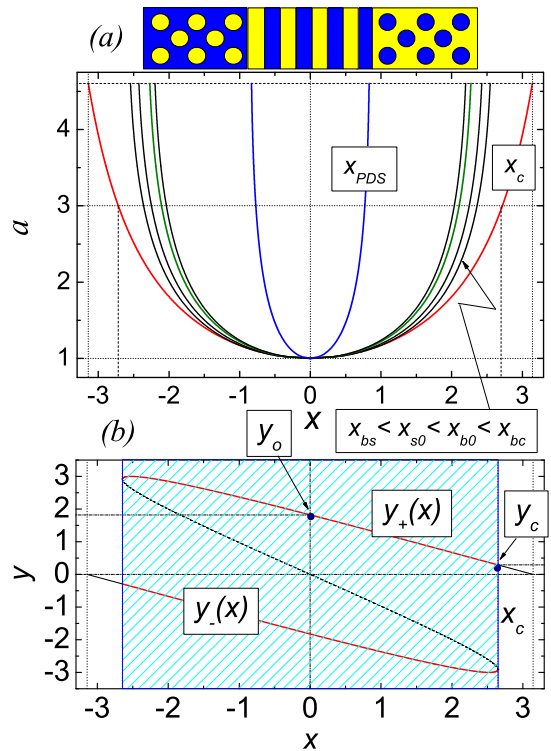}
\caption{ (color online) (a) Phase diagrams in $a-x$ plane for one
period of dHvA oscillations and (b) magnetic field dependencies of
the magnetization $y_{\pm}=y_{\pm}(x)$ (solid lines) calculated in
the framework of phase theory for $a=$3. The characteristic fields
$x_{PDS}, x_{bs}, x_{s0}, x_{b0}, x_{bc}$ and $x_{c}$ are explained
in the text. The sparse area in (b) corresponds to IDP. In
calculations of phase diagrams the value of $r_{c}/L=10^{-2}$ is
used.}\label{Phase Diagrams}
\end{figure}

The standard procedure of calculation of phase diagrams consists in
consideration of different configurations of domain patterns and
minimization of free energy density containing three terms, e. g.
dipole-dipole energy, the surface energy of the boundary states and
the energy of interaction with external magnetic field (Zeeman term)
\cite{Hubert}. In connection to this, there are two main mutually
related to each other features that make the calculations for
strongly correlated electron gas at the conditions of diamagnetic
instability different from analogous calculations in physics of
magnetic materials. The first, the magnetic moments of oppositely
magnetized uniform phases differ on values. And, the second, the
energy of interaction of magnetization with magnetic field is not
trivial, it is included explicitly in the first term of
Eq.~(\ref{eq:Free Energy Functional 2}) and defines the magnetic
field dependence of diamagnetic moments $y_{\pm}=y_{\pm}(x)$ of
uniformly magnetized phases within the period of dHvA oscillations
due to
\begin{equation} \label{eq: Uniform Magnetization}
y=-\partial_{x}G^{(unif)}=a \sin(x+y).
\end{equation}
We demonstrate these differences by consideration of isolated band
domains. Imbedding a uniformly magnetized band of width $W$ and of
magnetization $y_{-}-y_{+}$ into a plate of thickness $L$ and
opposite magnetization $y_{+}$, we can write the total energy
\begin{equation} \label{eq: Total Energy}
e^{(tot)}=\frac{g(w)}{2\pi}(\mu_{+}+\mu_{-})^{2}-\mu_{+}(\mu_{+}+\mu_{-})w+\lambda_{c},
\end{equation}
where the function
\begin{equation} \label{eq: Dipole-Dipole Energy}
g(w)=2w\tan^{-1}w+\frac{1}{2}\ln [w^{-2}(1+w^{-2})^{w^{2}-1}].
\end{equation}
is defined in \cite{Hubert} in slightly different form, $w=W/L$ is
reduced band width and $\mu_{\pm}=y_{+}(\pm x)$. In derivation of
Eq.~(\ref{eq: Total Energy}) we used $y_{-}(x)=-y_{+}(-x)$ (see,
also Fig.~\ref{Phase Diagrams}(b)). The first term in Eq.~(\ref{eq:
Total Energy}) is the self-energy of imbedded band, the second term
describes the interaction of the band with uniformly magnetized
matrix \cite{Hubert} and the third term corresponds to interface
boundary energy. Minimization of total energy of isolated band with
respect to $w$ results in
\begin{equation} \label{eq: Band Width}
\frac{2\pi\mu_{+}}{\mu_{+}+\mu_{-}}=\partial_{w}g,
\end{equation}
which defines the equilibrium band width $w=w(x,a)$ as a function of
magnetic field $x$ and differential magnetic susceptibility $a$.
Another equation corresponding to the saturation state is
$e^{(tot)}=0$ \cite{Hubert} which with the help of Eq.~(\ref{eq:
Band Width}) gives rise to
\begin{equation} \label{eq: Saturation}
\frac{4\pi\lambda_{c}}{(\mu_{+}+\mu_{-})^2}=w\partial_{w}g-g.
\end{equation}
For $a\gtrsim$2 the function $\mu_{\pm}$ can be approximated by
linear function $\mu_{\pm}=y_{0}\pm(y_{0}-y_{c})x/x_{c}$, where
$y_{c}=y(x_{c})$ is the value of magnetization at the bifurcation
set (see, Fig.~\ref{Phase Diagrams}(b)). In this case, instead of
Eqs~(\ref{eq: Band Width}) and (\ref{eq: Saturation}) we obtain
\begin{equation} \label{eq:Phase Boundary}
\left\{ \begin{array}{ll}
 1-x/\widetilde{x}_{c}=\pi[2\tan^{-1}w+w\ln(1+w^{-2})]\\
 2\pi \lambda_{c}/y^{2}_{0}=w^{2}\ln(1+w^{-2})+\ln(1+w^{2}),
  \end{array} \right.
\end{equation}
where $\widetilde{x}_{c}=(1-y_{c}/y_{0})^{-1}x_{c}\approx x_{c}$ is
a saturation field which in a limit $a>>$1 coincides with the
critical field $x_{c}$ defined by bifurcation set
(\ref{eq:Bifurcation Set}). Eqs.~(\ref{eq:Phase Boundary}) define a
parametric function $a=a(x_{s0})$ with $w$ as a parameter. This
function is plotted in Fig.~\ref{Phase Diagrams} together with the
family of phase boundary curves corresponding to other possible
pattern configurations, e. g. periodic parallel band structure,
bubble lattice and isolated bubbles, calculated in a similar way. We
adopted the indexes of characteristic fields \cite{Hubert}, e. g.
$x_{bc}$ is the bubble collapse field, $x_{b0}$ is the bubble
lattice saturation field, $x_{s0}$ is the saturation field for
isolated non-interacting bands and $x_{bs}$  is the strip-out
characteristic field. The magnetic field $x_{PDS}$ is the
characteristic field below which the PDS of parallel bands exists.
At the center of dHvA oscillations, $x=$0, the CDs of equal width
exist. With increasing of the small-scale field, $x$, the volume
fraction of up-domains increases in the expense of down-domains.
Close to the boundary $x \to x_{PDS}-0^{+}$, the distortions of
regular CD phase, e. g. bending, are expected. When
$x=x_{PDS}+0^{+}$, this structure transforms into the hexagonal
bubble lattice which is more favored energetically. The further
increase of the field $x$ results in the decrease of the bubble
density till the transformation of the lattice into the separated
bubbles which exist as a minority phase in the range
$x_{bs}<x<x_{bc}$. It follows from our analysis that the existence
of different inhomogeneous phases is strongly affected by
large-scale magnetic field $\mu_{0}H$, temperature $T$ and Dingle
temperature $T_{D}$ due to the corresponding dependencies of the
differential magnetic susceptibility $a=a(\mu_{0}H,T,T_{D})$ which
can be used for experimental studies of predicted effects.

\section{\label{sec:Conclusions}Conclusions}

We investigated theoretically the phase characteristics of the IDP
caused by the arising instability of strongly correlated electron
gas in high magnetic field and low temperature. We show that the
formation of IDP can realized through a variety of different
diamagnetic phases, including the plane-parallel band structure,
isolated non-interacting bands, hexagonal lattice of magnetic
bubbles and separated non-interacting bubbles.

The evolution of the domain patterns is different at the center of
dHvA period and away of it. While at $x=$0 the lowing of a
temperature results in formation of modulated structure in the
nearest vicinity of critical point which with further decrease of
temperature transforms into regular CD structure with the domains of
equal width, away of the center $x\ne$0 the system of strongly
correlated electron gas can underdo the series of phase transitions
with formation of different domain patterns depending on
differential magnetic susceptibility $a$.

We think that the appearance of distinct complex structure of IDT
has to affect the transport properties of electron gas under the
diamagnetic instability, but this is beyond the scope of the
article. The Condon domains provide an excellent system for
fundamental research, and we hope that our studies will stimulate
the further experiments on observation of electron instability at
the conditions of strong dHvA effect and investigation of intrinsic
structure of the non-uniform diamagnetic phases.

\begin{acknowledgments}
We are indebted to V. Egorov, R. Kramer and I. Sheikin for fruitful
discussions.
\end{acknowledgments}

\appendix*

\section{\label{sec:Cusp Catastrophe}Cusp Catastrophe}

\subsection{\label{Bifurcation Set}Bifurcation Set}

Let us introduce a new function
\begin{equation} \label{eq:phi}
\phi(z;\boldsymbol {\alpha})= z-\frac{\sin{z}}{1+\alpha_{1}}-\alpha_{2}~, \qquad \\
\end{equation}
which is a sufficiently smooth two-parametric function of
$(z,\boldsymbol {\alpha})\subset \mathbb{R}^{3}$, where $\boldsymbol
{\alpha}=(\alpha_{1}, \alpha_{2})$ is two-component vector which
controls the approach to critical point. The function
Eq.~(\ref{eq:phi}) describes the equilibrium properties of the
system of strongly correlated electron gas with magnetic induction
$b=z$ as the phase-state variable dependent on two adjustable, or
control parameters, e. g. reduced amplitude of dHvA oscillations $a$
and increment of internal magnetic field $x$, connected to the
components of the vector $\boldsymbol {\alpha}=(\alpha_{1},
\alpha_{2})$ by the relations
\begin{equation} \label{eq:alpha}
a=(1+\alpha_{1})^{-1}, \quad x=\alpha_{2}~. \\
\end{equation}
In mean-field theory of critical behavior of system \cite{Kadanoff},
the parameter $\alpha_{1}$ controls the amount of ordering, or the
value of order parameter (magnetic induction splitting in case of
DPT), while parameter $\alpha_{2}$ breaks the $Z(2)$ symmetry of
order parameter. We assume that the system has an equilibrium
$z=z_{0}$ with eigenvalue $\lambda=\phi_{z}(z_{0};\boldsymbol
{\alpha}_{0})=0$ at $\boldsymbol {\alpha}=\boldsymbol {\alpha}_{0}$,
e. g.
\begin{equation} \label{eq:double degenerate points}
\left\{ \begin{array}{ll}
 \phi(z;\boldsymbol {\alpha})=0\\
 \partial_{z}\phi(z;\boldsymbol {\alpha})=0.
  \end{array} \right.
\end{equation}

The Jacobian matrix of the system (\ref{eq:double degenerate
points})

\begin{equation} \label{eq:Jacobian}
\mathbf{J}= \left( \begin{array}{ccc}
\partial_{z}\phi & \partial_{\alpha_{1}}\phi & \partial_{\alpha_{2}}\phi \\
\partial_{zz}\phi & \partial_{z\alpha_{1}}\phi & \partial_{z\alpha_{2}}\phi
\end{array} \right)
\end{equation}
has a maximal rank $r=2$ at point $(z_{0}, \boldsymbol
{\alpha}_{0})$ due to non-zero value of the determinant

\begin{equation} \label{eq:rank}
\mathbf{det} \left( \begin{array}{ccc}
\partial_{\alpha_{1}}\phi & \partial_{\alpha_{2}}\phi \\
\partial_{z\alpha_{1}}\phi & \partial_{z\alpha_{2}}\phi
\end{array} \right) = \frac{1}{1+\alpha^{0}_{1}} \ne 0.
\end{equation}

Therefore, the system of Eqs.~(\ref{eq:double degenerate points})
defines a smooth curve $L$ in the joint space $\mathbb{R}^{3}$ of
phase-state and control variables ($z,\boldsymbol {\alpha}$).
According to Eqs.~(\ref{eq:double degenerate points}) this curve
corresponds to the double degenerated points, passes through the
point $(z_{0}, \boldsymbol {\alpha_{0}})$ belonging to the surface
of equilibrium states $\phi(z;\boldsymbol {\alpha})=0$ and can be
parameterized by $z$ due to non-zero value of the determinant
Eq.~(\ref{eq:rank}). Actually, in this case the existence of the two
following smooth functions of $z$
\begin{equation} \label{eq:bifurcation set 1}
 \boldsymbol {\alpha}=\boldsymbol {\alpha}(z),
\end{equation}
is provided by the implicit function theorem. For
$\phi(z;\boldsymbol {\alpha})$ defined by Eq.~(\ref{eq:phi}) one can
obtain

\begin{equation} \label{eq:bifurcation set 2}
\left\{ \begin{array}{ll}
 \alpha_{1}=-2 \sin^{2} (z/2)\\
 \alpha_{2}=z-\tan z.
  \end{array} \right.
\end{equation}
The standard projection operator map the curve $L\subset
\mathbb{R}^{3}$ onto the space of control variables $(\boldsymbol
{\alpha})$ giving rise to bifurcation set. The straightforward
calculation based on eliminating of $z$ in (\ref{eq:bifurcation set
2}) allows to obtain the following expression for the function
$\alpha_{2}=\alpha_{2}(\alpha_{1})$:
\begin{equation} \label{eq:bifurcation set 3}
 \alpha_{2}=\cos^{-1}(1+\alpha_{1})-\frac{(-\alpha_{1})^{1/2}(2+\alpha_{1})^{1/2}}{1+\alpha_{1}}.
\end{equation}
The function $\alpha_{2}=\alpha_{2}(\alpha_{1})$
Eq.~(\ref{eq:bifurcation set 3}) defines the bifurcation curve on
the plane of control variables $(\alpha_{1}, \alpha_{2})$ where a
fold bifurcation occurs. When $\alpha_{1}, \alpha_{2}<<$1,
Eq.~(\ref{eq:bifurcation set 3}) can be simplified resulting in
standard cusp bifurcation set
\begin{equation} \label{eq:bifurcation set 4}
 (2\alpha_{1})^{3}+(3\alpha_{2})^{2}=0,
\end{equation}
which is a semicubic parabola.

From Eq.~(\ref{eq:bifurcation set 3}) with use of relation
(\ref{eq:alpha}) recalling that for plate-like sample $x=x_{c}-y$
where $x_{c}$ is external magnetic field, we arrive to
Eq.~(\ref{eq:Bifurcation Set}).

In next section for completeness, we present the derivation of the
normal form for the system describe by Eq.~(\ref{eq:phi}). The real
use of the normal form consists in scaling the original problem to
the other one which is much simpler to solve.

\subsection{\label{Normal Form Derivation}Normal Form Derivation}

It follows from Eqs.~(\ref{eq:phi}), (\ref{eq:double degenerate
points}) that there exists the equilibrium point of the system $z=0$
for which the cusp bifurcation conditions
\begin{equation} \label{eq:cusp conditions}
\left\{ \begin{array}{ll}
 \partial_{z}\phi(0;\boldsymbol {0})=0\\
 \partial_{zz}\phi(0;\boldsymbol {0})=0
  \end{array} \right.
\end{equation}
are satisfied. It is important to note that
$\partial_{zzz}\phi(0;\boldsymbol {0}) \ne$0. Thus, expanding the
function $\phi(z;\boldsymbol {\alpha})$ Eq.~(\ref{eq:phi}) in a
Taylor series with respect to $z$ at $z=$0 and performing a linear
scaling of the phase-state variable $z$
\begin{equation} \label{eq:scaling}
 \eta = 6(1+\alpha_{1}) z, \qquad
\end{equation}
one can arrive to the following normal form for cusp bifurcation
\begin{equation} \label{eq:normal form}
 f(\eta;\boldsymbol {\beta})=\beta_{1}-\beta_{2}\eta+\eta^{3},
\end{equation}
where the components of the vector function $\boldsymbol
{\beta}=\boldsymbol {\beta}(\boldsymbol {\alpha})$ are the new
control variables defined by the following relations
\begin{equation} \label{eq:parameters}
 \beta_{1}=-6^{4}\alpha_{2}(1+\alpha_{1})^{4}, \quad \beta_{2}=-6^{3}\alpha_{1}(1+\alpha_{1})^{3}.
\end{equation}
The analysis of the normal form Eq.~(\ref{eq:normal form}) which is
topologically equivalent to the function Eq.~(\ref{eq:phi}) is
straightforward (see, e. g. \cite{Kuznetsov}). In particular, for
$\vert \alpha_{1} \vert<<$ 1 it gives the bifurcation set
(\ref{eq:bifurcation set 4}) in the vicinity of triple degenerate
point.

\end{document}